# Magnetic and microwave properties of $MFe_2O_4$-PZT (M =Ni, Co) and (La,Ca,Sr)$MnO_3$-PZT multilayer structures

A.A. Bush[1], Y.K. Fetisov[1], K.E. Kamentsev[1], V.F. Meshcheryakov[1], and G. Srinivasan[2]

[1]Moscow State Institute of radio Engineering, Electronics and Automation, 117454 Moscow, Russia
[2]Physics Department, Oakland University, Rochester, Michigan 48309-4401, USA

___________________________________________________________________________________

Abstract

Structural, magnetic and ferromagnetic resonance characterization studies have been performed on layered ferromagnetic-ferroelectric oxides that show strong magnetoelectric coupling. The samples contained thick films of ferrites or substituted lanthanum manganites for the ferromagnetic phase and lead zirconate titanate for the ferroelectric phase, and were sintered high temperatures. Results indicate defect free ferrites, but deterioration of manganite parameters due to diffusion at the interface and accounts for poor magnetoelectric coupling in manganite_PZT samples.

Keywords: Multilayer structures; Magnetic films; Ferromagnetic resonance

___________________________________________________________________________________

## Introduction

Multilayer structures consisting of alternate layers of magnetostrictive and piezoelectric materials show a strong magnetoelectric (ME) effect and are of interest for studies on ME interactions and for potential use in devices [1,2]. The ME effect in such a structure is a "product-property" and is proportional to the piezomagnetic and piezoelectric coefficients of the layers [3]. The purpose of present investigation is to determine the influence of the piezoelectric layers on magnetic and microwave properties of ferromagnetic layers in the layered composites.

## 1. Layered composites

Multilayer structures were fabricated from thick films synthesized by the doctor blade techniques. Nickel ferrite $NiFe_2O_4$ (NFO), cobalt ferrite $CoFe_2O_4$ (CFO), lanthanum-strontium manganite $La_{0.5}Sr_{0.5}MnO_3$ (LSM), or lanthanum-calcium manganite $La_{0.5}Ca_{0.5}MnO_3$ (LCM) was used as ferromagnetic layers. Lead zirconate-titanate $PbZr_{0.52}Ti_{0.48}O_3$ (PZT) was used as piezoelectric layer. Dense, well-bound multilayer structures with 0.03-0.05 cm in thickness and 1-5 $cm^2$ in area were fabricated. For all the structures thickness of a magnetic layer was equal to thickness of a piezoelectric layer. Samples studied and important parameters are summarized in Table 1.

Table 1. Parameters of the multilayer structures.

| Sample | Structure composition (number of layers), layer thickness |
|---|---|
| NFO-PZT | Nickel ferrite(15)–PZT(14), 14 $\mu m$ |
| CFO-PZT | Cobalt ferrite(10)-PZT(9), 26 $\mu m$ |
| LSMO-PZT | Lanthanum strontium manganite(21)- PZT(20), 10 $\mu m$ |
| LCMO-PZT | Lanthanum calcium manganite(9)- PZT(8), 35 $\mu m$ |

Results of X-ray studies of the composite structures showed that structural parameters of the composites were in good agreement with parameters for the constituent phases and no any new phases were formed at the interfaces because of diffusion.

## 2. Static magnetization

Measurements of static magnetization for all the samples were carried out at room temperature using a vibrating sample magnetometer in fields $H$ up to 6 kOe. Magnetization of the samples was then calculated taking into account vol-ume of the ferromagnetic phase. Figure1 shows typical magnetization curves for parallel ($H_{//}$) and perpendicular ($H_\perp$) orientations of external magnetic field with respect to the sample plane. Magnetization curves for the samples with different number of layers and layer thickness overlapped each other.

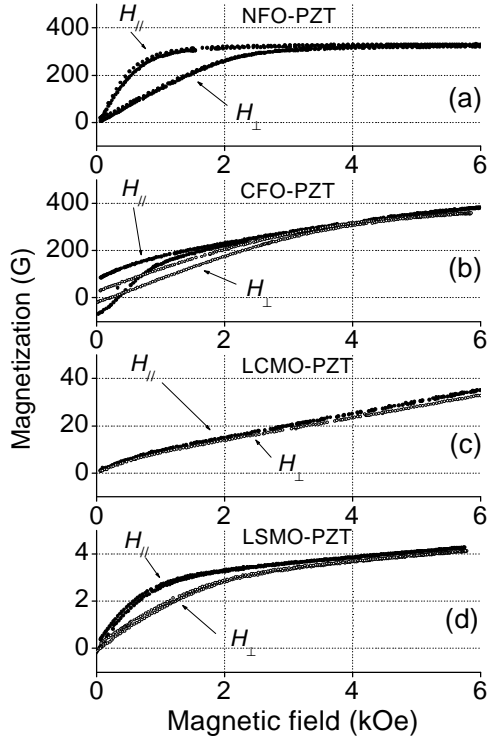

Fig.1 Magnetization for the multilayer structures at room temperature.

One observes for the NFO-PZT sample in Fig.1(a) a strong dependence of $M$ on the orientation of $H$. For in-plane magnetization, the saturation takes place at field about of 2 kOe, which is smaller than the saturation field for transverse magnetization. The saturation magnetic moment $M_s$ per unit volume of the NFO phase was 330 G, in agreement with the magnetization of bulk nickel ferrite [4]. No hysteresis was observed for the samples. For CFO-PZT [Fig.1(b)], there is significant dependence of $M$ on $H$ orientation as expected. Besides, a well-defined hysteresis is observed. Saturation of the sample magnetization occurs for $H$ higher than 6 kOe, where the magnetic moment equals 400 G, in agreement with reported values of 425 G [4]. Such an $M$ versus $H$ behavior is indicative of the expected large magnetic anisotropy for cobalt ferrite. Thus the magnetization curves of Fig.1 for ferrite-PZT samples correspond to magnetization for the bulk ferromagnetic counterparts. The key inference is that ferrites and PZT cosinter well and the PZT layers have no influence on magnetic properties of the structures.

For manganite-PZT structures, however, evidence for diffusion related degradation of the sample parameters was found. Bulk samples of the $La_{0.7}Ca_{0.3}MnO_3$ and $La_{0.7}Sr_{0.3}MnO_3$ order ferromagnetically at 260 K and 385 K, respectively [5]. The composites, however, showed a ferromagnetic $T_c$ of 240 K for LCMO-PZT and 301 K for LSMO-PZT. At room temperature, $M$ is on the order of 40 G for the LCMO-PZT sample [Fig. 1(c)] and no hysteresis was observed. The magnetization curves for H orientations overlapped each other since the sample is paramagnetic at room temperature. Results in Fig.1(d) for LSMO-PZT are somewhat different. The magnetization is relatively small and there is a small difference in magnetization curves corresponding to different orientations of $H$. There is clear evidence for ferromagnetic ordering at room temperature and a minor hysteresis is observed in Fig.1(d). The saturation magnetization is about of 4 G, well below expected values [6]. These reductions in $M$ and $T_c$ agree with our earlier observation of deterioration of magnetic parameters for when the layer thickness is reduced to 10 µm or the sintering temperature is increased over 1000 K. Thus the interface diffusion is a serious problem that needs to be resolved in manganite-PZT structures.

## 3. Ferromagnetic resonance

Ferromagnetic resonance (FMR) in the composite structures was investigated at room temperature at 9.8 GHz. Data were obtained for parallel ($H_{//}$) and perpendicular ($H_\perp$) orientations of the static field with respect to the sample plane. Figure 2 shows representative resonance spectra for all the samples.

The observed resonance field for $H_{//}$ is relatively low for NFO-PZT and is quite high for LCMO-PZT as expected due to a large magnetization for NFO-PZT compared to LCMO-PZT. The resonance field for $H_{//}$ for LSMO-PZT is surprisingly low although the magnetization is small and is indicative of a very high growth induced in-plane magnetic anisotropy in the sample. The FMR resonance line width is also strongly influenced by the anisotropy of the material. It is seen in Fig.2 that CFO-PZT has a very wide and irregular absorption line, while NFO-PZT has rather narrow and symmetrical line. This correlates with the hysteresis and coercivity in the magnetization for CFO-PZT and the absence of it for NFO-PZT.

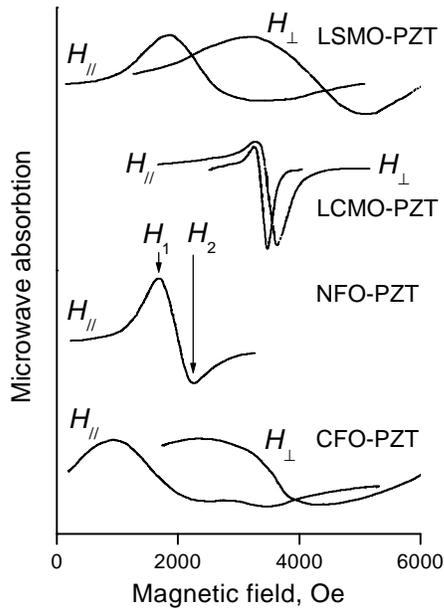

Fig.2: FMR spectra for the samples

The data on the resonance fields were then used to estimate magnetic parameters for the multilayer composites, such as the g-value, the effective magnetization $4\pi M_{\mathrm{eff}}=4\pi M-H_{\mathrm{A}}$, and the anisotropy field $H_{\mathrm{A}}$. Some of these parameters are listed in Table 2. For NFO-PZT, the g-value and the anisotropy field are in excellent agreement with values for bulk NFO samples [4]. Thus the two phases in NFO-PZT cosinter very well with practically no diffusion across the interface. For CFO-PZT layered samples, the g-value is higher than 2.7 for bulk CFO, but the anisotropy field agrees well with the expected value of 4.3 kOe [4]. It is therefore clear from magnetization and FMR studies that magnetic parameters for the ferrite phase in the layered samples were not affected by high temperature processing.

The samples with manganites, however, show evidence for diffusion of metal ions at the interface. In particular, FMR data reveal a large in-plane anisotropy for LSMO-PZT samples. Since bulk LSMO is magnetically isotropic at room temperature, $H_{\mathrm{A}}$ in the layered sample most likely originates from impurities and/or strain at the interface. The observation is in agreement with a reduction in both the ferromagnetic Curie temperature and magnetization for LSMO-PZT, as discussed earlier. The g-values for the layered samples are in agreement with values for bulk manganites [7].

Table 2. The FMR parameters of the structures.

| Sample | $\Delta H_{//}$, Oe | g | $4\pi M_{\mathrm{ef}}$, kGs | $H_{\mathrm{A}}$, kOe |
|---|---|---|---|---|
| NFO-PZT | 560 | 2.19 | 3.64 | 0.5 |
| CFO-PZT | 1800 | 3.27 | 1.34 | 3.7 |
| LCMO-PZT | 230 | 2.07 | 0.1 | 0.15 |
| LSMO-PZT | 1350 | 2.25 | 1.2 | -1.1 |

**Acknowledgments**
The research is supported by the US National Science Foundation (grant #DMR-0072144) and the Ministry of Education of Russia (grant #    ).